\documentclass[fp]{jpsj3}
\usepackage{txfonts}
\usepackage{graphicx}
\usepackage[usenames]{color}
\usepackage{amssymb}
\usepackage{amsmath}
\usepackage{bm}

\title{Anisotropic Charge Distribution Induced by Spin Polarization 
in La$_{0.6}$Sr$_{0.4}$MnO$_{3}$ Thin Films 
Studied by X-ray Magnetic Linear Dichroism}

\author{Goro Shibata$^1$\thanks{shibata@wyvern.phys.s.u-tokyo.ac.jp}, 
Kohei Yoshimatsu$^{1,2}$, 
Keisuke Ishigami$^1$, 
Takayuki Harano$^1$, 
Yukio Takahashi$^1$, 
Shoya Sakamoto$^1$, 
Yosuke Nonaka$^1$, 
Toshiharu Kadono$^1$, 
Mitsuho Furuse$^3$, 
Schuichiro Fuchino$^3$, 
Makoto Okano$^3$, 
Jun-ichi Fujihira$^4$, 
Akira Uchida$^4$, 
Kazunori Watanabe$^4$, 
Hideyuki Fujihira$^4$, 
Seiichi Fujihira$^4$, 
Arata Tanaka$^5$, 
Hiroshi Kumigashira$^2$, 
Tsuneharu Koide$^2$, 
and 
Atsushi Fujimori$^1$
}
\inst{
$^1$Department of Physics, University of Tokyo, Bunkyo-ku, Tokyo 113-0033, Japan \\
$^2$Photon Factory, Institute of Materials Structure Science, High 
Energy Accelerator Research Organization (KEK), Tsukuba, Ibaraki 305-0801, 
Japan \\
$^3$National Institute of Advanced Industrial Science and Technology 
(AIST), Tsukuba, Ibaraki 305-8568, Japan \\
$^4$Fujihira Co., Ltd., Tsukuba, Ibaraki 305-0047, Japan \\
$^5$Department of Quantum Matter, Graduate School of Advanced Sciences of Matter, 
Hiroshima University, Higashi-Hiroshima, Hiroshima 739-8530, Japan
} 

\abst{
Magnetic anisotropy of epitaxially grown thin films 
is affected by the strain from the substrates 
due to a combined effect of distorted electronic structure and 
spin-orbit interaction (SOI). 
As an inverse process, one expects an anisotropy of 
the electronic structure induced by magnetization in the presence of SOI. 
We have studied the charge-density anisotropy induced by 
magnetization in thin films of the ferromagnetic metal La$_{1-x}$Sr$_{x}$MnO$_3$ 
via x-ray magnetic linear dichroism (XMLD). 
XMLD measurements on thin films with various thicknesses have shown that 
the XMLD intensity is proportional to the square of the ferromagnetic moment. 
Using the XMLD sum rule and cluster-model calculation, 
it has been shown that 
more Mn 3$d$ electrons are distributed in orbitals elongated 
along the direction parallel to the spin polarization 
than in orbitals elongated in the direction perpendicular to it. 
The cluster-model calculation has shown that the effect of tensile strain 
from the SrTiO$_3$ substrate on the XMLD spectra is also consistent with the 
observed XMLD spectral line shapes. 
}

\begin{document}
\maketitle

\section{Introduction}

Magnetic anisotropy of ferromagnetic thin films and multilayers 
has been intensively studied so far 
both from technological and scientific interests \cite{Coey}.
From the technological point of view, 
materials with large magnetic anisotropy are desired 
for the realization of permanent magnets with higher coercive fields, 
magnetic recording media with higher density, 
and various spintronics devices. 
From the scientific point of view, 
clarifying the microscopic origin of magnetic anisotropy has been a 
challenging problem. 
Although it has been well established that magnetic anisotropy arises as a 
combined effect of the anisotropy of electronic states and spin-orbit interaction (SOI), 
consensus has not been reached yet regarding the detailed mechanism for it. 
In his seminal paper, Bruno \cite{Bruno} has shown by the perturbative treatment of SOI that the 
magnetocrystalline anisotropy (MCA) energy 
is proportional to the anisotropy of the orbital magnetic moment ($M_{\text{orb}}$), 
suggesting that the orbital-moment anisotropy (OMA) 
is the primary origin for MCA. 
Wang \textit{et al.} \cite{Wang_MCA_PRB93} 
have extended Bruno's theory, by incorpolating the `spin-flip' term 
in addition to the `spin-conservation' term, that 
the anisotropy of spin-density distribution, 
represented by `magnetic dipole' $M_\text{T}$ \cite{TXMCD_Stohr, SpinSum}, 
also contributes to the MCA \cite{vanderLaanJPCM98}. 
The theories by Bruno and Wang \textit{et al.} have been summarized 
by van der Laan \cite{vanderLaanJPCM98} in a more concise form. 
Several theoretical studies have predicted that the magnetic dipole $M_\text{T}$ 
may make a larger contribution to the MCA energy than the OMA 
\cite{FePt_Solovyev, FePt_Ravindran_calc, FeNi_MiuraShirai}. 
It has also been shown by several x-ray magnetic circular dichroism (XMCD) studies 
that the anisotropy of spin-density distribution $M_{\text{T}}$ 
is associated with MCA \cite{AngleDep_shibata, Miwa_Ncom17}. 
These studies suggests that the charge-density anisotropy 
can also affect the preferential orientation of 
spin magnetic moments. 

Since the electron orbitals and spins are coupled with each other through SOI, 
it is also expected, as an inverse process of the abovementioned process leading to MCA, that 
one can magnetically induce the OMA and 
anisotropic charge distribution by aligning the electron spin magnetic moments. 
X-ray linear dichroism (XLD) in core-level x-ray absorption spectroscopy (XAS) is a spectroscopic method 
which can probe the anisotropic charge distribution 
by measuring the differences in the XAS spectra between the two orthogonal linear polarizations. 
It is particularly called x-ray magnetic linear dichroism (XMLD) 
if the anisotropic charge distribution is magnetically induced. 
XMLD has been utilized for various magnetic thin films and multilayers 
in order to clarify the relationship between the electron spins and orbitals 
\cite{XMLD_Kuiper_Fe3O4_JElec97, XMLD_metals_PRB98, XMLD_Dhesi_Co_PRL01, XMLD_Kunes_PRB03, XMLD_Fe3O4_PRB06, XMLD_NiO_PRL07, XMLD_CoOx_PRB08, XMLD_EuO_PRL08}, 
especially for systems which exhibit perpendicular magnetic anisotropy \cite{XMLD_Dhesi_Co_PRL01, XMLD_CoOx_PRB08} 
and exchange bias \cite{XMLD_NiO_PRL07}. 
As for the thin films of ferromagnetic manganites such as La$_{1-x}$Sr$_{x}$MnO$_3$ (LSMO), 
the anisotropy of charge distribution 
between the out-of-plane and in-plane directions has been investigated 
via XLD \cite{TebanoPRB06, TebanoPRL08, ArutaPRB09, PesqueraNatCom12} and XMLD \cite{ArutaPRB09}. 
However, the charge anisotropy within the film plane 
has not been investigated in these studies. 

In the present article, we shall discuss 
intercoupling between the electron spins and charge anisotropy 
in LSMO thin films by the XMLD method. 
The relationship between XMLD and ferromagnetic moment is confirmed by 
the thickness dependence of XMLD.

\section{Methods}
LSMO ($x=0.4$) thin films were grown on SrTiO$_3$ (STO) (001) substrates 
by the laser molecular beam epitaxy method \cite{LaserMBE}. 
Due to the difference in the lattice constants between 
LSMO ($a=0.384\ \text{nm}$) and STO ($a=0.3905\ \text{nm}$), 
the LSMO films undergo tensile strain from the STO substrates \cite{Konishi}. 
The growth conditions of the films were essentially the same as 
that described in Ref.\ \citen{LSMO_Shibata}. 
The thicknesses of the LSMO films were between 2 and 15 unit cell (UC). 
The films were capped with 1 UC of La$_{0.6}$Sr$_{0.4}$TiO$_3$ and 
subsequently-deposited 2 UC of STO 
(see Fig.\ 1(a) in Ref.\ \citen{LSMO_Shibata}). 
The films were annealed in O$_2$ atmosphere in order to fill the oxygen vacancies 
after the deposition. 

\begin{figure}[tbp]
\centering
\includegraphics[width=0.9\linewidth]{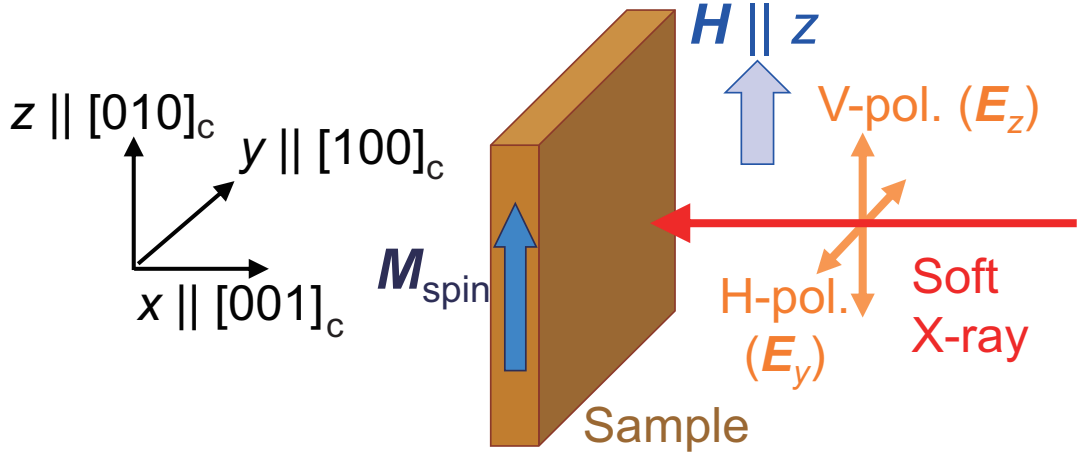} 
\caption{(Color Online) Schematic description of the experimental geometry of 
the present x-ray magnetic linear dichroism (XMLD) study. }
\label{XMLDgeometry}
\end{figure}

Figure \ref{XMLDgeometry} schematically describes the experimental geometry of XMLD. 
The x rays were incident on the sample normal to the film ([001]$_\text{c}$ direction, 
where the subscript `c' denotes that the indices are written 
in terms of the pseudocubic unit cell). 
The polarization of the x rays was either horizontal (denoted by H) or vertical (denoted by V). 
The films were mounted on the sample holder so that the in-plane [100]$_\text{c}$ direction 
was horizontal and the [010]$_\text{c}$ direction was vertical. 
A magnetic field $\bm{H}$ of $\mu _0 H = 0.1\ \text{T}$ 
(which was enough to saturate the magnetization along the 
in-plane direction) was applied along the vertical ([010]$_\text{c}$) direction, 
using the vector-XMCD apparatus \cite{Vector_Furuse, AngleDep_shibata}. 
This means that the V and H polarizations correspond to 
the electic field vector of the x rays ($\bm{E}$) 
parallel and perpendicular to the spin magnetic moment $\bm{M}_{\text{spin}}$, 
respectively. 
In the present article, 
the in-plane [010]$_\text{c}$ direction (i.e., the direction of the $\bm{M}_{\text{spin}}$) 
is chosen to be the $z$-axis, 
following the definition in Refs.\ \citen{XMLD_Thole85, XMLD_Carra_PhysicaB93}. 
The $x$-axis is chosen to be the sample-normal ([001]$_\text{c}$) direction 
(i.e., opposite to the wavevector of the incident x rays) 
and $y$-axis to be the [100]$_\text{c}$ direction (i.e., the in-plane direction 
perpendicular to $\bm{M}_{\text{spin}}$). 
We define XMLD as $\text{XAS (V)}-\text{XAS (H)}$ here. 

The XMLD experiments were performed at the beamline BL-16A2 of KEK Photon Factory (KEK-PF) 
installed with a twin Apple II-type undulator. 
The measurement temperature $T$ was $30\ \text{K}$. 
The spectra were measured in the total electron-yield (TEY) mode. 
The base pressure of the measurement chamber was $\sim 1 \times 10^{-9}$ Torr. 
The obtained XAS and XMLD spectra were analyzed with the cluster-model calculation 
using the `Xtls' code (version 8.5) \cite{TanakaCluster}. 
Details of the calculation methods are described 
in the previous report \cite{AngleDep_shibata}. 

\section{Results and Discussion}
\label{sec:result_xmld}

\begin{figure}[tbp]
\centering
\includegraphics[width=0.6\linewidth]{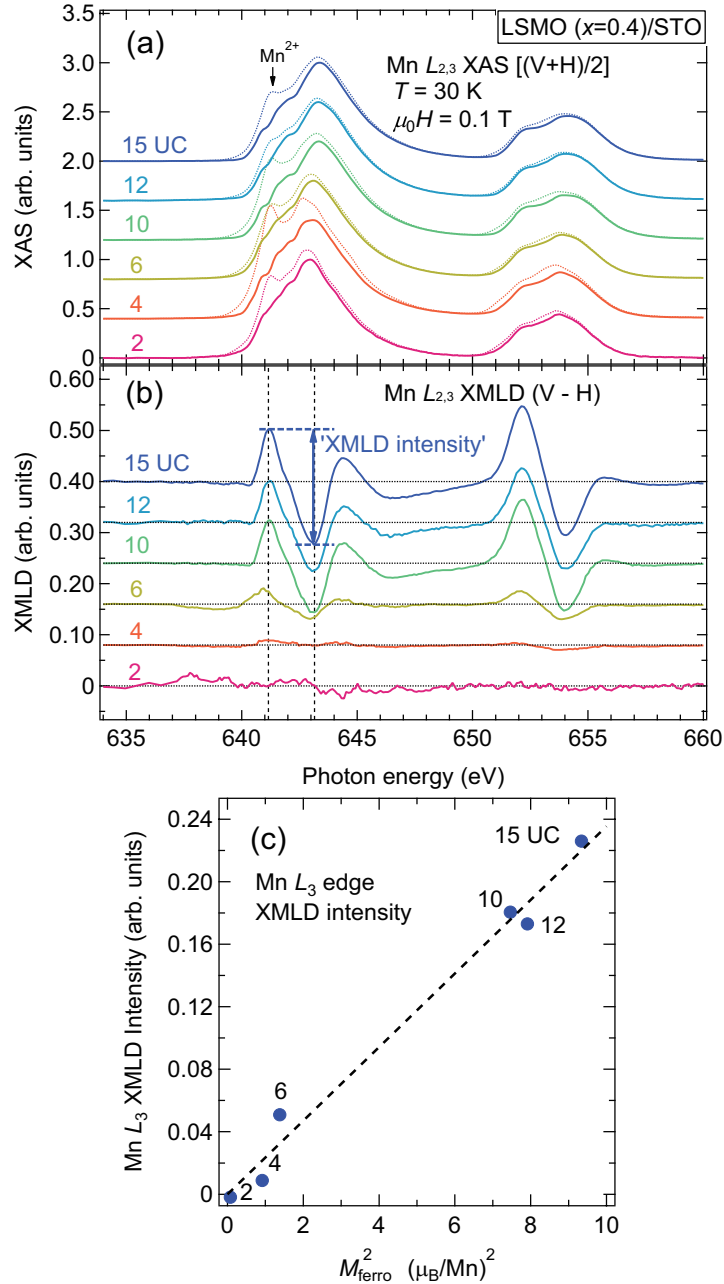} 
\caption{(Color Online) 
XMLD on La$_{1-x}$Sr$_{x}$MnO$_3$ (LSMO) thin films 
grown on SrTiO$_3$ (STO) (001) substrates 
with various thicknesses. 
(a) X-ray absorption spectroscopy (XAS) spectra at the Mn $L_{2,3}$ absorption edge 
averaged over both the polarizations. 
Dashed curves are the raw spectra and solid curves are the spectra after 
subtracting the extrinsic signals of Mn$^{2+}$ (see text). 
(b) XMLD spectra at the Mn-$L_{2,3}$ absorption edge measured at $\mu _0 H = 0.1\ \text{T}$. 
XMLD is defined as XAS(V)$-$XAS(H), where V and H, respectively, denote the 
vertical and horizontal linear polarization 
(i.e., $\bm{E} \parallel \bm{M}_{\text{spin}}$ and $\bm{E} \perp \bm{M}_{\text{spin}}$). 
The definition of the XMLD intensity is shown by an arrow. 
(c) XMLD intensities of LSMO/STO thin films with various thicknesses 
plotted as a function of 
ferromagnetic moment $M_{\text{ferro}}^2$ deduced from 
x-ray magnetic circular dichroism (XMCD) \cite{LSMO_Shibata}. 
Dashed line shows the result of a least-square fitting with a linear function, 
in which the constant term has been fixed to zero. 
}
\label{XMLDspectra}
\end{figure}

Figure \ref{XMLDspectra}(a) shows the Mn $L_{2,3}$-edge XAS spectra 
of the LSMO thin films 
averaged over both the H and V polarizations. 
In the raw spectra, XAS signals originating from Mn$^{2+}$ overlap, 
as shown by dotted curves in Fig.\ \ref{XMLDspectra}(a). 
Such XAS signals of Mn$^{2+}$ are sometimes observed in manganite thin films 
due to extrinsic effects such as oxygen reduction at the surface \cite{Mn2plus}. 
We have, therefore, subtracted these extrinsic Mn$^{2+}$ signals 
following the method presented in the previous study \cite{Takeda_GaMnAs}. 
This affects the absolute values of the XMLD intensities by $\sim 10\%$ at most, 
but does not change our main conclusion. 
After having subtracted the Mn$^{2+}$ signals, the spectral line shapes are almost identical to 
those in previous studies \cite{LSMO_Koide, TebanoPRB06, TebanoPRL08, ArutaPRB09, PesqueraNatCom12, LSMO_Shibata, AngleDep_shibata}. 
The peak positions of the spectra are shifted to lower photon energies 
with decreasing thickness, 
indicating that the valence of Mn gradually decreases. 
This is the same tendency as the previous XMCD study, 
which may be due to electron doping at the interfaces from the substrates 
and/or from oxygen vacancies \cite{LSMO_Shibata}. 
Figure \ref{XMLDspectra}(b) shows the XMLD spectra 
of the LSMO thin films with various thicknesses. 
We note that the lower signal-to-noise ratio of the XMLD spectrum for the 2-UC film 
than those for the other films is due to the smaller photocurrent intensities 
for the 2-UC film caused by the small sample volume and 
the high resistivity of the thinnest film \cite{LSMO_Huijben}. 
The XMLD intensity gradually decreases 
as the thickness of LSMO is reduced, 
while the spectral line shape of XMLD is essentially unchanged. 
Here, the XMLD intensity is defined as the difference between the signal intensities 
at 641.1 eV and at 643.1 eV 
(which are the peak and the dip positions of the XMLD spectra 
for the 15 UC film, respectively), as shown in Fig.\ \ref{XMLDspectra}(b). 
Note that this definition has been adopted because it is less affected by 
the procedure of the background subtraction 
compared to the spectral areas of XMLD. 
In Fig.\ \ref{XMLDspectra}(c), thus estimated XMLD intensities 
are plotted against the square of the ferromagnetic (FM) moment $M_{\text{ferro}}^2$, 
which has been estimated from the magnetization curves measured by XMCD 
\cite{LSMO_Shibata}. 
The plot clearly shows that the XMLD intensity is 
proportional to $M_{\text{ferro}}^2$. 
In general, the XMLD intensity is proportional to the square of 
the local spin magnetic moment \cite{XMLD_Thole85}. 
The above result suggests that the 
XMLD signals originate from the FM phases rather than the antiferromagnetic (AFM) phases, 
which was possibly present in the sample as an impurity phase.

\begin{figure}[tbp]
\centering
\includegraphics[width=0.7\linewidth]{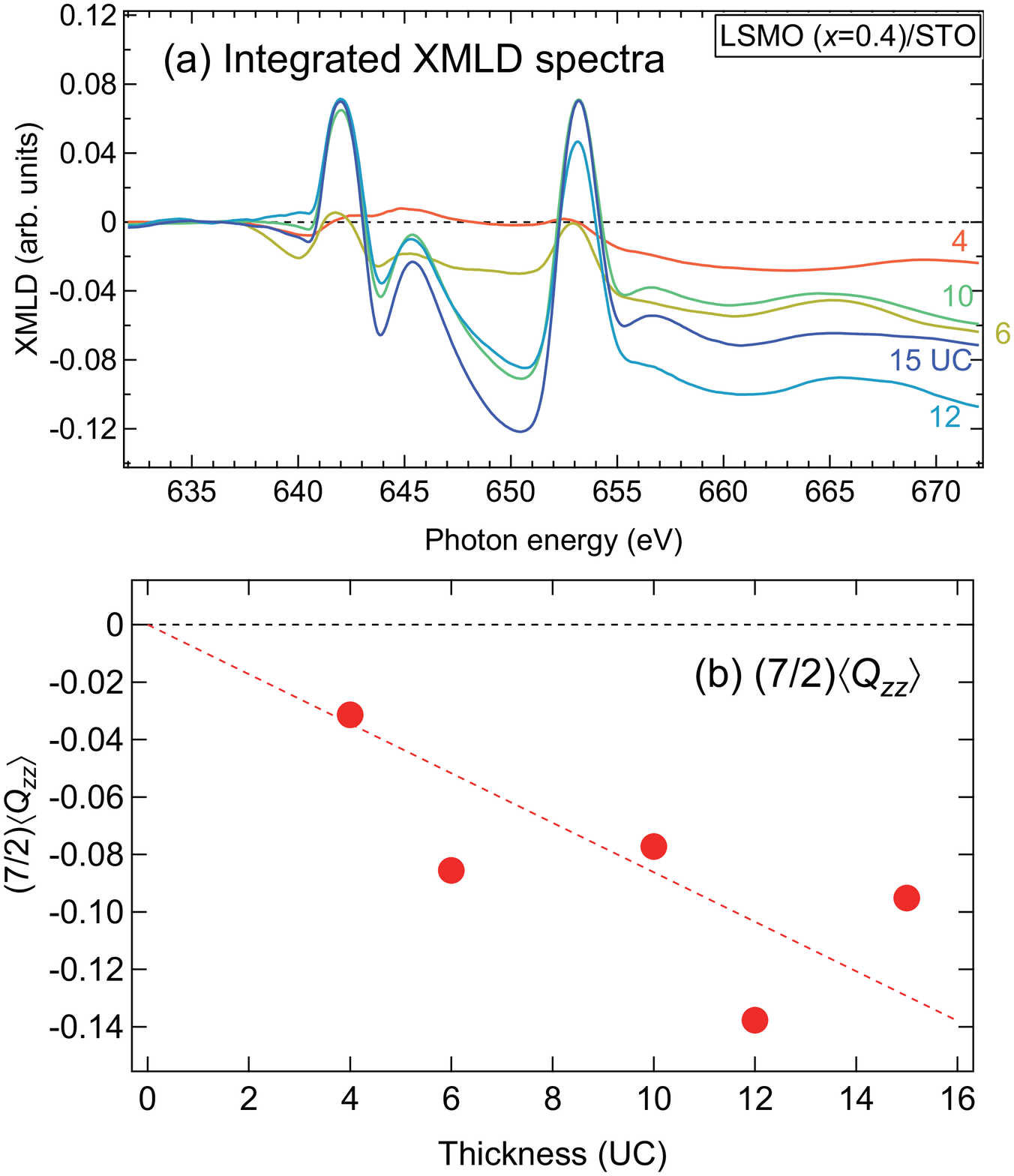} 
\caption{(Color Online) Estimation of charge-distribution anisotropy 
from the integrals of XMLD. 
(a) Integrals of the XMLD spectra shown in Fig.\ \ref{XMLDspectra}(b). 
(b) Electric quadrupole moment $(7/2)\langle Q_{zz} \rangle$ deduced from the 
integral of the XMLD spectra over the Mn $L_{2,3}$ edge. 
Data for 2 UC have been omitted because the XMLD signal 
was within the noise level. 
Dashed line shows the result of a least-square fitting with a linear function, 
in which the constant term has been fixed to zero. 
}
\label{XMLDinteg}
\end{figure}

According to the sum rule for XLD, \cite{XMLD_Carra_PhysicaB93} 
the integral of the XMLD spectra over the $L_3$ and $L_2$ absorption edges 
gives the electric quadrupole moment $\langle Q_{zz} \rangle = \langle 1 - 3z^2/r^2 \rangle$, 
namely, the anisotropy of the charge distribution. 
Figure \ref{XMLDinteg}(a) shows the integrated XMLD spectra calculated from Fig.\ \ref{XMLDspectra}(b), 
and the values of $(7/2)\langle Q_{zz} \rangle$ deduced from them are shown 
in Fig. \ref{XMLDinteg}(b) as a function of thickness. 
We note that $(7/2)\langle Q_{zz} \rangle$ is equal to $+2$ ($-2$) 
for the $d_{x^2-y^2}$ ($d_{3z^2-r^2}$) orbital \cite{TXMCD_Stohr}. 
The values of $\langle Q_{zz} \rangle$ are negative for all the thicknesses 
and the absolute values of $\langle Q_{zz} \rangle$ gradually increases with thickness, 
as the films turn from the paramagnetic into the ferromagnetic states. 
Without the spin magnetic moment $\bm{M}_{\text{spin}}$, 
the electron distribution should be isotropic between the in-plane $y$- and $z$-directions 
because the films have a tetragonal crystal symmetry. 
The negative values of $\langle Q_{zz} \rangle$ ($= \langle 1 - 3z^2/r^2 \rangle$) 
shows that the 
electrons are more densely distributed along the $z$-directions than the $y$-direction 
due to the presence of $\bm{M}_{\text{spin}}$, namely, 
the electron orbitals are `elongated' along the spin direction through SOI. 
The observed charge-density anisotropy $(7/2)\langle Q_{zz} \rangle \sim -0.1$ 
corresponds to the preferential occupation of the $d_{3z^2-r^2}$ orbital by $\sim 10\%$ 
compared to the $d_{x^2-y^2}$ orbital. 

\begin{figure}[t]
\centering
\includegraphics[width=0.7\linewidth]{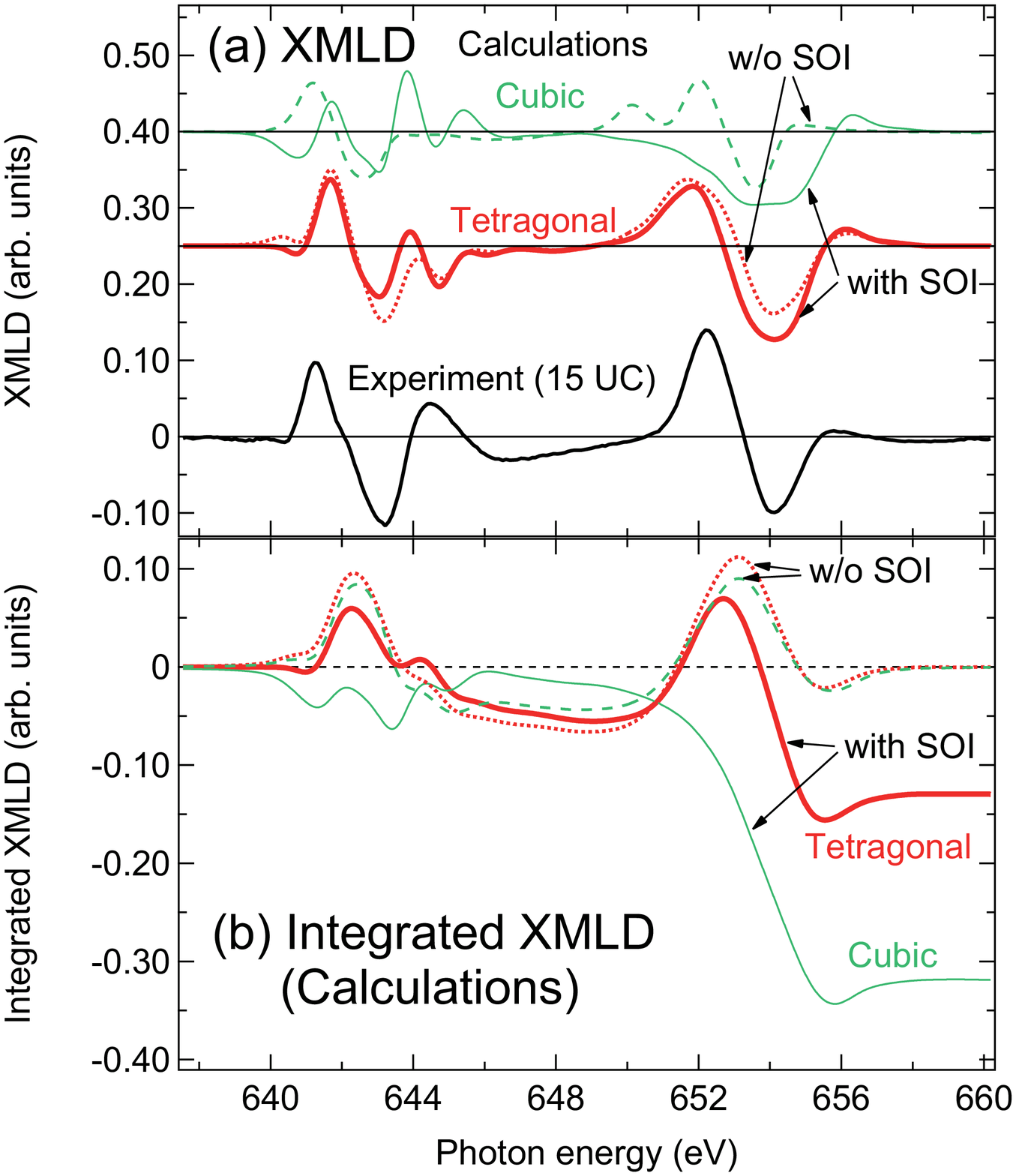} 
\caption{(Color Online) (a) Simulated XMLD spectra using the MnO$_6$ cluster model. 
(b) Integrals of the simulated XMLD spectra in (a). 
In both panels, thin (green) and thick (red) solid curves show the simulated spectra 
with Mn 3$d$ spin-orbit interaction (SOI) 
assuming the octahedral ($O_h$) and tetragonally ($D_{4h}$)-tensiled 
crystal symmetry, respectively. 
Dashed (green) and dotted (red) curves are the simulated spectra without Mn 3$d$ SOI 
for the $O_h$ and $D_{4h}$ symmetry, respectively. 
Black solid curve at the bottom of Panel (a) shows 
the experimental XMLD spectrum for the LSMO thin film of 15 UC thickness. 
}
\label{ClusterXMLD}
\end{figure}

In Fig.\ \ref{ClusterXMLD}(a), the experimentally obtained XMLD spectrum (for 15 UC) 
is compared with the theoretical ones calculated using the MnO$_6$ cluster model 
with octahedral ($O_h$) or tetragonal ($D_{4h}$) symmetry 
with respect to the out-of-plane ($x$) axis. 
For the simulation with $D_{4h}$ symmetry, the in-plane $y^2-z^2$ orbital 
has a lower energy than the out-of-plane $3x^2-r^2$ orbital by $0.08\ \text{eV}$ 
in order to simulate the tensile epitaxial strain. 
The parameter values used for the simulations are chosen to be 
the same as those in Ref.\ \citen{AngleDep_shibata}. 
In order to see the effects of SOI, the XMLD spectra 
in both the cases with and without SOI of Mn $3d$ electrons 
have been calculated. 
Comparing the calculated spectra with the experimental one, 
it can be seen that the calculation with $D_{4h}$ symmetry under tensile strain 
is closer to the experimental spectra than that with $O_{h}$ symmetry, 
especially for the spectral line shapes at the Mn $L_2$ edge. 
This is consistent with the tensile epitaxial strain from the STO substrate. 
Furthermore, as shown in Fig.\ \ref{ClusterXMLD}(b), 
the integrals of the simulated XMLD spectra over the Mn $L_3$ and $L_2$ edges 
become negative in the presence of SOI of Mn $3d$ electrons, 
while they are equal to zero in the absence of SOI for both the crystal symmetries. 
The negative XMLD integrals suggest that $\langle Q_{zz} \rangle < 0$, 
namely, the charge density is higher along the $z$-direction than the $y$-direction, 
which is again consistent with the experiment. 
These simulations also demonstrate that SOI of Mn 3$d$ electrons 
is indispensable for the experimentally observed 
charge-density anisotropy within the plane. 

The present result illustrates that the charge distribution of Mn $3d$ electrons 
is elongated along the spin direction. 
This means that the state in which the electron orbitals are elongated 
along the spins is more energetically favorable. 
This observation may be related to the strain-induced magnetic anisotropy 
in manganite thin films which has been previously reported \cite{Tsui, KwonJMMM97}. 
It is known that the magnetic easy directions of LSMO thin films 
are out-of-plane in the case of compressive strain 
and in-plane in the case of tensile strain \cite{Tsui, KwonJMMM97}. 
The charge density of spin-polarized Mn $3d$ electrons 
under the compressive and tensile strain 
is shown to be higher along the out-of-plane and in-plane directions, respectively 
\cite{Konishi, AngleDep_shibata}. 
Therefore, it follows that LSMO thin films tend to be magnetized 
parallel to the directions along which the Mn $3d$ orbitals are elongated, 
the same tendency as the present XMLD result. 
The present study corroborates that the strain-induced magnetic anisotropy in 
LSMO thin films originates from the combining effect of SOI and 
the charge-density anisotropy of spin-polarized Mn $3d$ electrons. 
The magnetic anisotropy 
and the charge anisotropy are related with each other in two ways: 
One is that the charge anisotropy of spin-polarized electrons 
yields finite magnetic dipole $M_\text{T}$, 
resulting in finite MCA energy through the spin-flip term 
in perturbation theory with respect to SOI by Wang \textit{et al.} 
\cite{Wang_MCA_PRB93, vanderLaanJPCM98}. 
The other is that the OMA contributes to the MCA energy 
through Bruno's spin-conservation term \cite{Bruno, vanderLaanJPCM98} 
and that the observed charge anisotropy is indirectly induced by the OMA. 
In order to see which of the magnetic dipole $M_\text{T}$ and the OMA 
plays a more important role in the MCA of LSMO thin films, 
further experimental and theoretical studies are needed. 

\section{Conclusion}
\label{sec:conc_xmld}

We have studied the magnetically-induced anisotropic charge distribution 
in the LSMO ($x=0.4$)/STO (001) thin films via XMLD. 
From the thickness dependence of the XMLD spectra, it has been shown that 
the XMLD intensity is proportional to the square of the FM moment of the film, 
suggesting that the XMLD signals originate from the FM phase in the LSMO thin films 
rather than the possible AFM impurity phase. 
The electric quadrupolar moment $\langle Q_{zz} \rangle$, 
which represents the anisotropy of the charge distribution, 
is shown to be negative, indicating that 
the in-plane ($y^2-z^2$) orbital of the Mn 3$d$ electrons 
is slightly elongated along the direction of the spins. 
This is consistent with the theoretical prediction 
based on the MnO$_6$ cluster model with tensile strain and with SOI. 
The observed charge anisotropy, i.e., the elongation of the electron orbital 
along the spin direction, 
may be related to the strain-induced magnetic anisotropy in LSMO thin films.

\begin{acknowledgments}


We would like to thank Kenta Amemiya and Masako Sakamaki 
for valuable technical support at KEK-PF. 
We would also like to thank Hiroki Wadati for providing us with information about 
the XLD studies of LSMO thin films. 
This work was supported by a Grant-in-Aid for Scientific Research from 
the JSPS (22224005, 15H02109, 15K17696, and 16H02115). 
The experiment was done under the approval of the Photon Factory 
Program Advisory Committee 
(proposal No.\ 2016S2-005, No.\ 2013S2-004, No.\ 2016G066, No.\ 2014G177, No.\ 2012G667, and 2015S2-005). 
G.S.\ acknowledges support from 
Advanced Leading Graduate Course for Photon Science (ALPS) 
at the University of Tokyo and the JSPS Research Fellowships 
for Young Scientists (Project No. 26.11615). 
A.F.\ is an adjunct member of Center for Spintronics Research Network (CSRN), 
the University of Tokyo, under Spintronics Research Network of Japan (Spin-RNJ). 

\end{acknowledgments}

\bibliography{68911.bib}

\begin{thebibliography}{10}

\bibitem{Coey}
J.~M.~D. Coey: {\em Magnetism and Magnetic Materials} (Cambridge University
  Press, New York, 2009).

\bibitem{Bruno}
P.~Bruno: Phys. Rev. B {\bfseries 39} (1989) 865.

\bibitem{Wang_MCA_PRB93}
D.-s. Wang, R.~Wu, and A.~J. Freeman: Phys. Rev. B {\bfseries 47} (1993) 14932.

\bibitem{TXMCD_Stohr}
J.~St\"ohr and H.~K\"onig: Phys. Rev. Lett. {\bfseries 75} (1995) 3748.

\bibitem{SpinSum}
P.~Carra, B.~T. Thole, M.~Altarelli, and X.~Wang: Phys. Rev. Lett. {\bfseries
  70} (1993) 694.

\bibitem{vanderLaanJPCM98}
G.~van~der Laan: J. Phys. Condens. Matter {\bfseries 10} (1998) 3239.

\bibitem{FePt_Solovyev}
I.~V. Solovyev, P.~H. Dederichs, and I.~Mertig: Phys. Rev. B {\bfseries 52}
  (1995) 13419.

\bibitem{FePt_Ravindran_calc}
P.~Ravindran, A.~Kjekshus, H.~Fjellv\aa{}g, P.~James, L.~Nordstr\"om,
  B.~Johansson, and O.~Eriksson: Phys. Rev. B {\bfseries 63} (2001) 144409.

\bibitem{FeNi_MiuraShirai}
Y.~Miura, S.~Ozaki, Y.~Kuwahara, M.~Tsujikawa, K.~Abe, and M.~Shirai: J. Phys.:
  Condens. Matter {\bfseries 25} (2013) 106005.

\bibitem{AngleDep_shibata}
G.~Shibata, M.~Kitamura, M.~Minohara, K.~Yoshimatsu, T.~Kadono, K.~Ishigami,
  T.~Harano, Y.~Takahashi, S.~Sakamoto, Y.~Nonaka, K.~Ikeda, Z.~Chi, M.~Furuse,
  S.~Fuchino, M.~Okano, J.-i. Fujihira, A.~Uchida, K.~Watanabe, H.~Fujihira,
  S.~Fujihira, A.~Tanaka, H.~Kumigashira, T.~Koide, and A.~Fujimori: npj
  Quantum Mater. {\bfseries 3} (2018) 3.

\bibitem{Miwa_Ncom17}
S.~{Miwa}, M.~{Suzuki}, M.~{Tsujikawa}, K.~{Matsuda}, T.~{Nozaki}, K.~{Tanaka},
  T.~{Tsukahara}, K.~{Nawaoka}, M.~{Goto}, Y.~{Kotani}, T.~{Ohkubo},
  F.~{Bonell}, E.~{Tamura}, K.~{Hono}, T.~{Nakamura}, M.~{Shirai}, S.~{Yuasa},
  and Y.~{Suzuki}: Nat. Commun. {\bfseries 8} (2017) 15848.

\bibitem{XMLD_Kuiper_Fe3O4_JElec97}
P.~Kuiper, B.~Searle, L.-C. Duda, R.~Wolf, and P.~van~der Zaag: J. Electron.
  Spectrosc. Relat. Phenom. {\bfseries 86} (1997) 107 .

\bibitem{XMLD_metals_PRB98}
M.~M. Schwickert, G.~Y. Guo, M.~A. Tomaz, W.~L. O'Brien, and G.~R. Harp: Phys.
  Rev. B {\bfseries 58} (1998) R4289.

\bibitem{XMLD_Dhesi_Co_PRL01}
S.~S. Dhesi, G.~van~der Laan, E.~Dudzik, and A.~B. Shick: Phys. Rev. Lett.
  {\bfseries 87} (2001) 067201.

\bibitem{XMLD_Kunes_PRB03}
J.~Kune\ifmmode~\check{s}\else \v{s}\fi{} and P.~M. Oppeneer: Phys. Rev. B
  {\bfseries 67} (2003) 024431.

\bibitem{XMLD_Fe3O4_PRB06}
E.~Arenholz, G.~van~der Laan, R.~V. Chopdekar, and Y.~Suzuki: Phys. Rev. B
  {\bfseries 74} (2006) 094407.

\bibitem{XMLD_NiO_PRL07}
E.~Arenholz, G.~van~der Laan, R.~V. Chopdekar, and Y.~Suzuki: Phys. Rev. Lett.
  {\bfseries 98} (2007) 197201.

\bibitem{XMLD_CoOx_PRB08}
G.~van~der Laan, E.~Arenholz, R.~V. Chopdekar, and Y.~Suzuki: Phys. Rev. B
  {\bfseries 77} (2008) 064407.

\bibitem{XMLD_EuO_PRL08}
G.~van~der Laan, E.~Arenholz, A.~Schmehl, and D.~G. Schlom: Phys. Rev. Lett.
  {\bfseries 100} (2008) 067403.

\bibitem{TebanoPRB06}
A.~Tebano, C.~Aruta, P.~G. Medaglia, F.~Tozzi, G.~Balestrino, A.~A. Sidorenko,
  G.~Allodi, R.~De~Renzi, G.~Ghiringhelli, C.~Dallera, L.~Braicovich, and N.~B.
  Brookes: Phys. Rev. B {\bfseries 74} (2006) 245116.

\bibitem{TebanoPRL08}
A.~Tebano, C.~Aruta, S.~Sanna, P.~G. Medaglia, G.~Balestrino, A.~A. Sidorenko,
  R.~De~Renzi, G.~Ghiringhelli, L.~Braicovich, V.~Bisogni, and N.~B. Brookes:
  Phys. Rev. Lett. {\bfseries 100} (2008) 137401.

\bibitem{ArutaPRB09}
C.~Aruta, G.~Ghiringhelli, V.~Bisogni, L.~Braicovich, N.~B. Brookes, A.~Tebano,
  and G.~Balestrino: Phys. Rev. B {\bfseries 80} (2009) 014431.

\bibitem{PesqueraNatCom12}
D.~Pesquera, G.~Herranz, A.~Barla, E.~Pellegrin, F.~Bondino, E.~Magnano,
  F.~S\'{a}nchez, and J.~Fontcuberta: Nat. Commun. {\bfseries 3} (2012) 1189.

\bibitem{LaserMBE}
K.~Horiba, H.~Ohguchi, H.~Kumigashira, M.~Oshima, K.~Ono, N.~Nakagawa,
  M.~Lippmaa, M.~Kawasaki, and H.~Koinuma: Rev. Sci. Instrum. {\bfseries 74}
  (2003) 3406.

\bibitem{Konishi}
Y.~Konishi, Z.~Fang, M.~Izumi, T.~Manako, M.~Kasai, H.~Kuwahara, M.~Kawasaki,
  K.~Terakura, and Y.~Tokura: J. Phys. Soc. Jpn. {\bfseries 68} (1999) 3790.

\bibitem{LSMO_Shibata}
G.~Shibata, K.~Yoshimatsu, E.~Sakai, V.~R. Singh, V.~K. Verma, K.~Ishigami,
  T.~Harano, T.~Kadono, Y.~Takeda, T.~Okane, Y.~Saitoh, H.~Yamagami, A.~Sawa,
  H.~Kumigashira, M.~Oshima, T.~Koide, and A.~Fujimori: Phys. Rev. B {\bfseries
  89} (2014) 235123.

\bibitem{Vector_Furuse}
M.~Furuse, M.~Okano, S.~Fuchino, A.~Uchida, J.~Fujihira, S.~Fujihira,
  T.~Kadono, A.~Fujimori, and T.~Koide: IEEE Trans. Appl. Supercond. {\bfseries
  23} (2013) 4100704.

\bibitem{XMLD_Thole85}
B.~T. Thole, G.~van~der Laan, and G.~A. Sawatzky: Phys. Rev. Lett. {\bfseries
  55} (1985) 2086.

\bibitem{XMLD_Carra_PhysicaB93}
P.~Carra, H.~K\"onig, B.~Thole, and M.~Altarelli: Physica B {\bfseries 192}
  (1993) 182 .

\bibitem{TanakaCluster}
A.~Tanaka and T.~Jo: J. Phys. Soc. Jpn. {\bfseries 63} (1994) 2788.

\bibitem{Mn2plus}
M.~P. de~Jong, I.~Bergenti, V.~A. Dediu, M.~Fahlman, M.~Marsi, and C.~Taliani:
  Phys. Rev. B {\bfseries 71} (2005) 014434.

\bibitem{Takeda_GaMnAs}
Y.~Takeda, M.~Kobayashi, T.~Okane, T.~Ohkochi, J.~Okamoto, Y.~Saitoh,
  K.~Kobayashi, H.~Yamagami, A.~Fujimori, A.~Tanaka, J.~Okabayashi, M.~Oshima,
  S.~Ohya, P.~N. Hai, and M.~Tanaka: Phys. Rev. Lett. {\bfseries 100} (2008)
  247202.

\bibitem{LSMO_Koide}
T.~Koide, H.~Miyauchi, J.~Okamoto, T.~Shidara, T.~Sekine, T.~Saitoh,
  A.~Fujimori, H.~Fukutani, M.~Takano, and Y.~Takeda: Phys. Rev. Lett.
  {\bfseries 87} (2001) 246404.

\bibitem{LSMO_Huijben}
M.~Huijben, L.~W. Martin, Y.-H. Chu, M.~B. Holcomb, P.~Yu, G.~Rijnders,
  D.~H.~A. Blank, and R.~Ramesh: Phys. Rev. B {\bfseries 78} (2008) 094413.

\bibitem{Tsui}
F.~Tsui, M.~C. Smoak, T.~K. Nath, and C.~B. Eom: Appl. Phys. Lett. {\bfseries
  76} (2000) 2421.

\bibitem{KwonJMMM97}
C.~Kwon, M.~Robson, K.-C. Kim, J.~Gu, S.~Lofland, S.~Bhagat, Z.~Trajanovic,
  M.~Rajeswari, T.~Venkatesan, A.~Kratz, R.~Gomez, and R.~Ramesh: J. Magn.
  Magn. Mater. {\bfseries 172} (1997) 229 .

\end{thebibliography}

\end{document}